\documentclass[sigconf]{acmart}

\usepackage{comment}
\usepackage{threeparttable}

\newcommand{\equalcontribmark}{\texorpdfstring{\textsuperscript{\textdagger}}{}}

\newcommand{\correspondingauthormark}{\texorpdfstring{\textsuperscript{*}}{}}

\AtBeginDocument{%
  }

\setcopyright{rightsretained}
\makeatletter
\renewcommand{\@copyrightowner}{UT-Battelle, LLC}
\renewcommand{\@copyrightpermission}{%
This manuscript has been authored in part by UT-Battelle, LLC, under contract
DE-AC05-00OR22725 with the U.S. Department of Energy (DOE). The U.S. Government
retains and the publisher, by accepting the article for publication, acknowledges that
the U.S. Government retains a non-exclusive, paid-up, irrevocable, worldwide license
to publish or reproduce the published form of this manuscript, or allow others to do so,
for U.S. Government purposes. DOE will provide public access to these results of
federally sponsored research in accordance with the DOE Public Access Plan
(\url{https://energy.gov/doe-public-access-plan}).%
}
\makeatother
\copyrightyear{2026}
\acmYear{2026}
\acmDOI{XXXXXXX.XXXXXXX}

\acmConference[UrbanAI '26]
{The 4th ACM SIGSPATIAL International Workshop on Advances in Urban AI}
{November 3, 2026}
{Riverside, CA, USA}

\acmBooktitle{UrbanAI '26: The 4th ACM SIGSPATIAL International Workshop on Advances in Urban AI, November 3, 2026, Riverside, CA, USA}
\acmISBN{978-1-4503-XXXX-X/2026/11}

\begin{document}

% ----------------------------------------------------------------
% TITLE AND AUTHORS
% ----------------------------------------------------------------
\title{Adapting a Large Language Model Crash-Severity Pipeline to Tennessee: Performance Across Sampling Strategies}

\author{Abhilasha Saroj\equalcontribmark\correspondingauthormark}
\email{sarojaj@ornl.gov}
\affiliation{%
  \institution{Oak Ridge National Laboratory}
  \city{Oak Ridge}
  \state{Tennessee}
  \country{USA}
}

\author{Pranav Govindu\equalcontribmark}
\email{pgovindu@vols.utk.edu}
\affiliation{%
  \institution{University of Tennessee at Knoxville}
  \city{Knoxville}
  \state{Tennessee}
  \country{USA}
}

\author{Bharat Sharma}
\email{sharmabd@ornl.gov}
\affiliation{%
  \institution{Oak Ridge National Laboratory}
  \city{Oak Ridge}
  \state{Tennessee}
  \country{USA}
}

\author{Usman Ahmed}
\email{uahmed@utk.edu}
\affiliation{%
  \institution{University of Tennessee - Oak Ridge Innovation Institute}
  \city{Oak Ridge}
  \state{Tennessee}
  \country{USA}
}

% Copy the following block for each additional author.
% \author{Second Author}
% \email{second.author@example.com}
% \affiliation{%
%   \institution{Institution Name}
%   \city{City}
%   \state{State}
%   \country{Country}
% }

% Copy the following block for each additional author.
% \author{Second Author}
% \email{second.author@example.com}
% \affiliation{%
%   \institution{Institution Name}
%   \city{City}
%   \state{State}
%   \country{Country}
% }

% Copy the following block for each additional author.
% \author{Second Author}
% \email{second.author@example.com}
% \affiliation{%
%   \institution{Institution Name}
%   \city{City}
%   \state{State}
%   \country{Country}
% }

\renewcommand{\shortauthors}{Saroj, Govindu, Sharma, Ahmed}

\begin{abstract}
State crash databases differ in structure, coding, and injury-severity distributions, limiting direct reuse of predictive workflows across jurisdictions. This study adapts the SafeTraffic Copilot large language model (LLM) crash-severity workflow to a three-year Tennessee inventory of 624,392 crashes. Tennessee crash, roadway, vehicle, and person attributes were harmonized and converted into textual prompts while unavailable values were preserved rather than inferred. Llama 3.1 8B was fine-tuned using low-rank adaptation to classify five injury-severity categories. Random, county-based, and severity-balanced sampling strategies were evaluated using separate in-sample and unseen-test experiments; unseen-test experiments used 70/15/15 training, validation, and test splits. On their respective unseen test sets, random and county-based models achieved weighted F1-scores near 80\%, but macro F1 remained below 47\% and fatal-crash F1 below 27\%, showing that strong aggregate performance can mask weak recognition of rare outcomes. Within its balanced evaluation population, the severity-balanced model achieved weighted and macro F1-scores of 57.7\% and a fatal-crash F1-score of 65.9\%, yielding more even class-level performance. Because each sampling strategy used a different test subset, cross-strategy differences are descriptive rather than controlled rankings. The results highlight the importance of interpreting LLM crash-severity performance together with sampling design, class balance, class-level metrics, and evaluation-population composition.
\end{abstract}

\maketitle

\begingroup
  \renewcommand{\thefootnote}{\textdagger}
  \footnotetext{These authors contributed equally to this work and share first authorship.}
\endgroup

\begingroup
  \renewcommand{\thefootnote}{*}
  \footnotetext{Corresponding author: Abhilasha Saroj (sarojaj@ornl.gov).}
\endgroup

% ----------------------------------------------------------------
% MAIN PAPER
% ----------------------------------------------------------------
\section{Introduction}
Traffic crash records combine event, roadway, environmental, vehicle, and person information. These records support injury-severity analysis, but their relational structure, missing values, unobserved heterogeneity, and strong concentration in less severe outcomes make them difficult to model reliably \cite{savolainen2011statistical,mannering2016heterogeneity}. Reviews of machine-learning severity models identify class imbalance, transferability, evaluation, and interpretation as persistent challenges \cite{wen2021mlreview}.

Differences among state reporting systems create an additional barrier to model reuse. The Model Minimum Uniform Crash Criteria (MMUCC) provides voluntary national guidance for crash, vehicle, person, and roadway data elements, yet states retain jurisdiction-specific fields, definitions, categorical codes, and reporting practices \cite{nhtsa2024mmucc}. A model developed for one jurisdiction therefore faces both schema differences and distribution shift when applied elsewhere.

Most statistical and machine-learning methods convert a crash into a fixed vector of tabular variables. Large language models (LLMs) offer an alternative by representing heterogeneous attributes as a coherent textual description. This representation can preserve relationships among crash-level, vehicle-level, and person-level information within a common input structure, although it does not remove the need for careful data harmonization or class-sensitive evaluation.

Recent studies apply LLMs to crash forecasting, severity inference, and text-based crash reasoning \cite{dezarza2023llm,zhen2024llmcrash,fan2024learning}. SafeTraffic Copilot converts multi-source crash records into general, infrastructure, event, and unit prompt sections and fine-tunes Llama 3.1 models to generate constrained outcome tokens \cite{zhao2025safetraffic,grattafiori2024llama3}. It uses low-rank adaptation (LoRA) for parameter-efficient fine-tuning \cite{hu2022lora}.

This study adapts the SafeTraffic workflow to a three-year Tennessee crash inventory containing around 624,392 records. The adaptation requires state-specific field mapping, consistent treatment of unavailable values, and alignment of Tennessee severity labels with model output tokens. The experiments compare the observed aggregate and class-level performance profiles associated with random, county-based, and severity-balanced samples. The analysis distinguishes in-sample performance, calculated on records used for model fitting, from unseen test-data performance, calculated on records excluded from training and validation. Because each sampling strategy uses its own test subset, the comparisons are interpreted descriptively rather than as a controlled test of training-sample effects. 

\subsection{Objectives}
The primary objective of this study is to evaluate whether an LLM-based crash-severity pipeline can be reproduced and adapted to a large state-owned crash inventory while retaining the receiving jurisdiction's reported information. The study has three specific objectives: (1) reproduce the released SafeTraffic model-facing workflow; (2) develop a defensible Tennessee harmonization and textualization process; and (3) characterize aggregate and class-level results for random, county-based, and severity-balanced samples. The analysis examines whether strong overall performance is also reflected in the identification of rare serious-injury and fatal crashes. Although the sampled datasets are similar in size, each sampling strategy produces a different mix of crash records and a separate test set. Therefore, the results show how each sampling strategy performs but do not allow a direct comparison among the strategies.

\section{Literature Review}

\subsection{Crash Inventory Issues}

Police-reported crash inventories are primarily designed to document crashes and support statewide safety management rather than to provide analysis-ready inputs for a uniform modeling pipeline. A single crash may be represented through linked crash, roadway, vehicle, driver, and occupant records, with varying numbers of vehicles and persons associated with each event. Constructing one model input therefore requires explicit decisions regarding record linkage, repeated entities, aggregation, missing values, and categorical encoding \cite{nhtsa2024mmucc,zhao2025safetraffic}.

% MMUCC promotes greater consistency in crash reporting by defining recommended crash type, vehicle, person, and roadway data elements.
MMUCC promotes greater consistency in crash reporting by defining recommended data elements for the crash, the vehicle, the roadway, and the people involved, including their condition and state at the time of the crash. However, MMUCC adoption is voluntary, and states retain jurisdiction-specific fields, definitions, categorical codes, and reporting practices \cite{nhtsa2024mmucc}. Cross-state adaptation consequently requires semantic harmonization rather than simple column matching. Fields with similar meanings may use different names or coding systems, similarly named fields may have different definitions, and some source fields may have no defensible counterpart in the receiving jurisdiction.

Crash-severity inventories are also strongly imbalanced because fatal and serious-injury crashes are much less frequent than no-injury or lower-severity crashes. A model can therefore achieve high accuracy or weighted F1 while performing poorly on the classes of greatest safety interest \cite{wen2021mlreview,santos2022literature,kotsyubynska2026machine}. Macro F1, class-level precision and recall, and the number of test observations in each class are necessary to reveal this behavior. Sampling and evaluation design must also be considered together because a naturally distributed test set measures performance under the observed population distribution, whereas a severity-balanced test set provides a class-sensitive diagnostic evaluation but does not represent the natural statewide crash distribution.

\subsection{LLMs for Crash Severity Characterization}

Crash-severity research includes statistical approaches, such as discrete-choice and ordered-response models, and machine-learning approaches that learn nonlinear relationships from structured variables \cite{savolainen2011statistical,manneringbhat2014analytic,zhang2018comparing}. Statistical models provide explicit assumptions and interpretable parameter estimates, whereas machine-learning models can represent complex interactions with fewer prespecified functional forms. Neither family of methods automatically resolves class imbalance, missing information, distribution shift, or weaknesses in evaluation design \cite{wen2021mlreview,santos2022literature,kotsyubynska2026machine}. In addition, the use of highly flexible black-box models in safety applications requires careful consideration of interpretability and validation \cite{rudin2019stop}.

Most conventional crash-severity pipelines represent a crash using a fixed vector of structured variables. Original police-written narratives, when available, are commonly analyzed separately or reduced to manually defined categories. LLM-based textualization provides an alternative representation in which structured roadway, environmental, crash, vehicle, and person attributes can be converted into a common textual sequence \cite{fan2024learning,dezarza2023llm,zhao2025safetraffic}. This approach can preserve the relationships among multiple crash entities within a consistent model-facing format.

Textualization changes the form of feature engineering rather than eliminating it. The workflow still requires record linkage, schema mapping, categorical harmonization, treatment of unavailable values, prompt construction, and controls against target leakage. It is also important to distinguish original officer-written crash narratives from textual descriptions generated from structured database fields. The present study primarily concerns the latter: structured Tennessee crash records are mapped into consistent textual descriptions for model input.

LLMs also remain vulnerable to class imbalance, distribution shift, target leakage, and limited interpretability. Attribution methods can indicate which words, sentences, or input sections influence a model prediction, but they do not establish causal relationships between those factors and crash severity \cite{zhao2025safetraffic,rudin2019stop}. The value of an LLM-based workflow must therefore be demonstrated through transparent field mapping, leakage controls, class-sensitive evaluation, and comparison with appropriate statistical and machine-learning baselines. \textit{The present study focuses on workflow adaptation and sampling-sensitive performance rather than claiming that LLMs eliminate the limitations of conventional crash-severity models.}

\subsection{LLMs for Crash Inference}
Recent studies have examined the use of LLMs for crash forecasting, crash-severity inference, and reasoning from textualized crash information \cite{dezarza2023llm,zhen2024llmcrash,fan2024learning}. Prompt-based studies indicate that LLMs can map crash descriptions to a finite set of severity categories and that domain-informed prompts can improve alignment between generated outputs and predefined severity labels \cite{zhen2024llmcrash}. A related line of work treats crash records as a domain-specific language dataset. CrashLLM, for example, textualizes infrastructure, environmental, traffic, and crash information and fine-tuned LLMs to predict crash type, injury severity, and the number of injuries \cite{fan2024learning}.

SafeTraffic Copilot extends this approach through an artificial-intelligence-and-expert textualization workflow, LoRA-based parameter-efficient fine-tuning, finite severity-label tokens, token-probability-based confidence scores, and sentence-level attribution \cite{zhao2025safetraffic,hu2022lora}. Its input is organized into general, infrastructure, event, and unit information. The finite-label-token design adds predefined crash-outcome tokens to the model vocabulary and frames severity inference as next-token generation over those labels rather than as unrestricted text generation.

The interpretation of ``severity inference'' depends on the information provided to the model. When the input contains crash, vehicle, occupant, and other information collected during or after the event, the task is more accurately described as retrospective crash-severity classification. It should not be interpreted as prospective or early severity prediction unless fields that directly or indirectly reveal the final injury outcome are excluded. The Tennessee adaptation therefore preserves unavailable values without inference and requires an audit of model inputs for potential severity leakage.

\textbf{Although LLM-based crash-severity studies are expanding, evidence remains limited on the independent reproduction of a publicly released workflow and its adaptation to a substantially larger operational state inventory with different schemas, categorical codes, reporting practices, and severity distributions}. Sampling design introduces an additional challenge because it changes both the representation of rare outcomes during training and, when strategy-specific test subsets are used, the composition of the evaluation population. The Tennessee application addresses this practical gap by adapting the previously developed SafeTraffic LLM workflow to state-specific crash records and reporting aggregate, macro, class-level, and class-support results for random, county-based, and severity-balanced evaluation sets. 
% Because the strategies use different test subsets, the results characterize strategy-specific performance profiles rather than providing a controlled estimate of the effect of training-sample design.

\subsection{Paper Scope and Contributions}

\begin{itemize}
\item The study independently reproduces the publicly released SafeTraffic version 0.1.0 model-facing workflow and adapts it for application to a large, three-year Tennessee crash inventory \cite{zhao2025safetraffic,zhao2025safetrafficcode}.

\item It develops a state-specific harmonization and textualization process that maps Tennessee crash, roadway, vehicle, and person records into the SafeTraffic prompt structure while preserving reported information and representing unsupported values as unavailable rather than inferring them.

\item It establishes a sampling-sensitive evaluation framework that examines random, county-based, and severity-balanced datasets using aggregate, macro, and class-level measures while explicitly accounting for class support and differences among strategy-specific test populations.

\end{itemize}

\section{Model Transferability Framework}
\begin{figure*}[htbp]
    \centering
    \includegraphics[width=\textwidth]{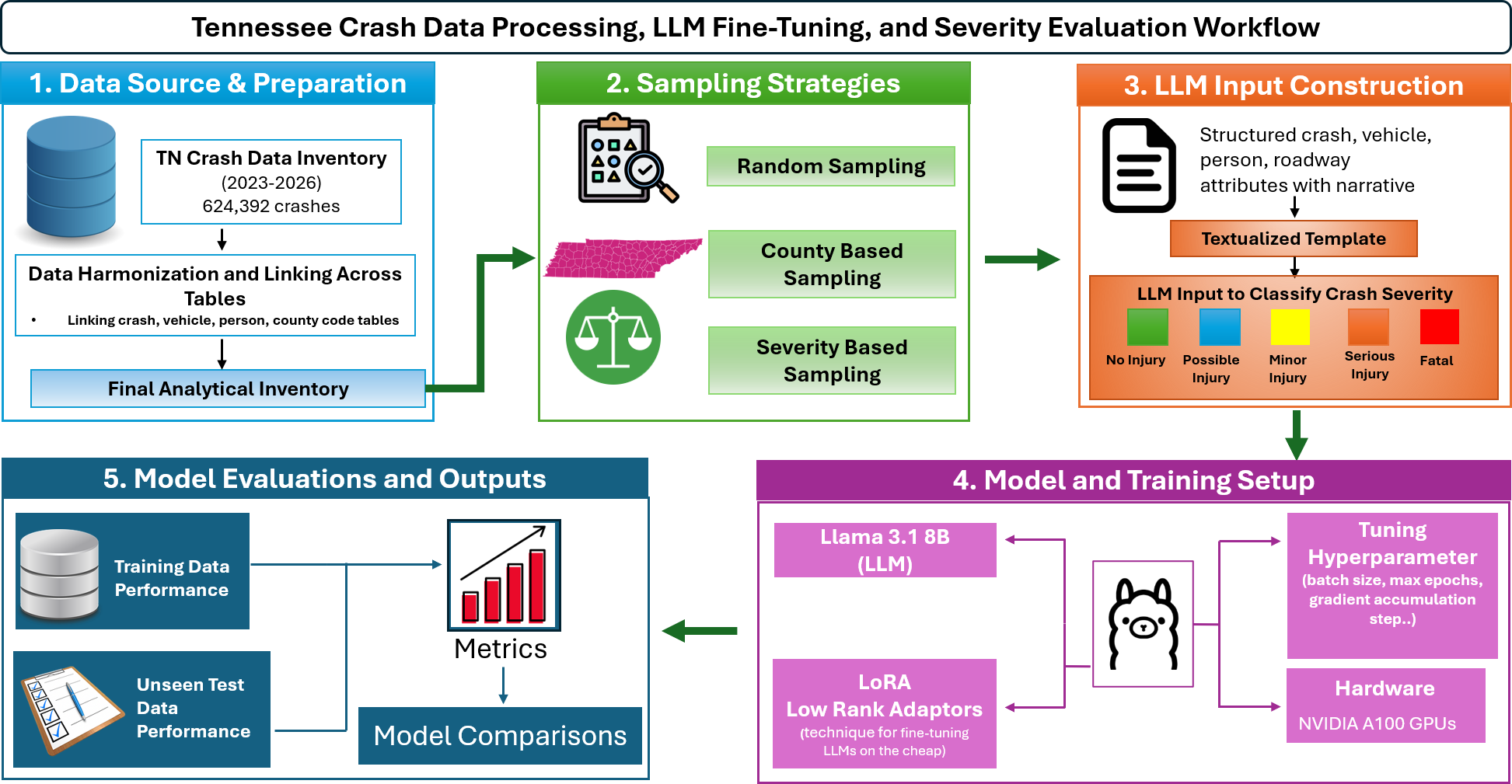}
    \caption{End-to-end transfer and fine-tuning workflow for Tennessee crash-severity inference.}
    \Description{A workflow diagram showing Tennessee crash-data preparation, field mapping, textualization, sampling, LLM fine-tuning, and performance evaluation.}
    \label{fig:training-workflow}
\end{figure*}

The workflow-adaptation framework separates data preparation from model training. First, Tennessee crash, roadway, vehicle, and person records are linked at the crash-event level. Second, state-specific fields and categorical values are mapped to the safety concepts used by the released pipeline. Third, the mapped content is organized into general, infrastructure, event, and unit prompt sections. Fourth, three sampling strategies create datasets emphasizing the observed statewide distribution, geographic representation, or severity balance. Fifth, each strategy-specific dataset is divided into training, validation, and test subsets and used to fine-tune the same LLM configuration. Finally, generated severity tokens are converted to Tennessee labels and evaluated using aggregate and class-level measures. Because the test subsets differ across strategies, the framework characterizes each strategy-specific performance profile but does not isolate the effect of training sampling on a common test population. Figure~\ref{fig:training-workflow} summarizes the process.

\subsection{Model Replication}

The replication follows the public SafeTraffic release tagged as version 0.1.0 \cite{zhao2025safetrafficcode}. The release provides state- and task-specific scripts and documents a Python 3.9 environment with Ubuntu 22.04, CUDA 12.1, and NVIDIA A100 graphics processing units (GPUs). Each instance contains a system prompt, a user prompt describing the crash, and a target prompt containing the expected special token. Consistent with the published method, LoRA adapters provide parameter-efficient fine-tuning rather than updating all base-model parameters \cite{hu2022lora}.

Replication is performed before Tennessee adaptation so that model implementation problems can be separated from data-mapping problems. The released Illinois and Washington files are used to verify reading, tokenization, training, and evaluation through the original interface. The Tennessee workflow then preserves the same model-facing format while replacing the source-specific preprocessing. The Illinois and Washington data descriptions below document the release subsets used for this verification; they are not presented as cross-state performance experiments in the current results.

\subsubsection{Model Data Description}

The SafeTraffic article reports cleaned 2022 datasets containing 42,715 Illinois crashes and 16,188 Washington crashes before construction of the broader SafeTraffic Event dataset \cite{zhao2025safetraffic}. The public replication files used in this study contain smaller subsets. Their sizes and distributions are reported below to distinguish the released replication inputs from the complete datasets described in the original article.

\noindent\textbf{Illinois Release Subset}

The Illinois release subset contains 14,814 crashes from 2022 and covers all 102 Illinois counties. The supplied partitions include 2,000 training records, 6,407 validation records, and 6,407 test records.

\begin{table}[h]
\centering
\caption{County and Injury-Severity Distribution of the Illinois Release Subset}
\label{tab:illinois-release}

\begin{tabular}[t]{|l|r|}
\hline
\textbf{County Name} & \textbf{Records} \\ \hline
Cook County          & 7,340 \\ \hline
Du Page              & 869   \\ \hline
Will                 & 734   \\ \hline
Lake                 & 639   \\ \hline
Kane                 & 538   \\ \hline
\end{tabular}
\quad
\begin{tabular}[t]{|l|r|r|}
\hline
\textbf{Severity} & \textbf{Count} & \textbf{Share} \\ \hline
No Apparent Injury & 5,193 & 35.05\% \\ \hline
Minor Injury       & 3,688 & 24.90\% \\ \hline
Possible Injury    & 3,649 & 24.63\% \\ \hline
Serious Injury     & 1,954 & 13.19\% \\ \hline
Fatal              & 330   & 2.23\%  \\ \hline
\end{tabular}

\end{table}

Table~\ref{tab:illinois-release} shows that the subset is geographically concentrated despite covering all counties. Cook County contributes nearly half of the records. Fatal crashes account for 2.23\% of the subset. The supplied partitions are retained only to verify the released workflow.

\noindent\textbf{Washington Release Subset}

The Washington release subset contains 4,000 crashes from 2022 and covers 34 of the state's 39 counties. The supplied partitions include 2,000 training records, 1,000 validation records, and 1,000 test records.

\begin{table}[h]
\centering
\caption{Injury-Severity Distribution of the Washington Release Subset}
\label{tab:washington-release}

\begin{tabular}[t]{|l|r|r|}
\hline
\textbf{Severity} & \textbf{Count} & \textbf{Share} \\ \hline
No Apparent Injury & 2,274 & 56.85\% \\ \hline
Minor Injury       & 995   & 24.88\% \\ \hline
Possible Injury    & 579   & 14.47\% \\ \hline
Serious Injury     & 118   & 2.95\%  \\ \hline
Fatal              & 34    & 0.85\%  \\ \hline
\end{tabular}

\end{table}

Table~\ref{tab:washington-release} shows that no apparent injury is the largest severity class, representing 56.85\% of the subset. Fatal crashes account for 0.85\%. These records provide the second source dataset used to verify the released workflow.

\subsubsection{Tennessee Data Description}

\begin{figure}[htbp]
    \centering
    \includegraphics[width=\linewidth]{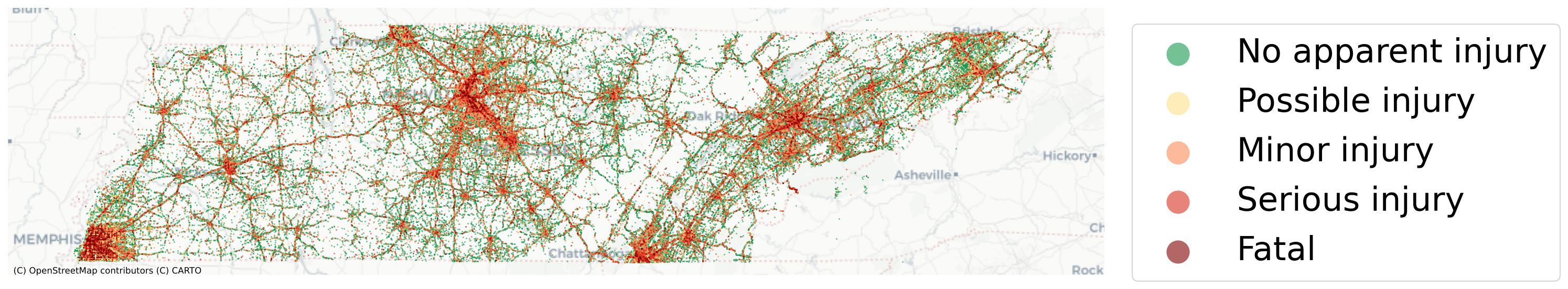}
    \caption{Spatial and temporal distribution of the three-year Tennessee crash inventory.}
    \Description{Maps and charts summarize the geographic distribution and yearly composition of Tennessee crash records in the three-year inventory.}
    \label{fig:database-distribution}
\end{figure}

The Tennessee inventory contains 624,392 crash records collected over three years. It is substantially larger and temporally broader than the release subsets, making it suitable for evaluating adaptation to an operational state database. Figure~\ref{fig:database-distribution} presents the spatial and temporal composition of the inventory, and Table~\ref{tab:injury_severity_new} reports the five injury-severity categories.

A state-specific preprocessing layer maps and organizes Tennessee field names and categorical codes to the corresponding SafeTraffic concepts such as general, event, and unit (vehicle specifications) sections. Values without a direct, defensible counterpart are represented consistently as unknown or unavailable. The target labels remain no apparent injury, possible injury, minor injury, serious injury, and fatal.

The inventory is strongly imbalanced: no apparent injury represents 78.54\% of records, whereas serious injury and fatal crashes represent 2.21\% and 0.53\%, respectively. Consequently, accuracy and weighted F1 are interpreted together with macro and class-level F1. This imbalance also motivates the direct comparison of random, county-based, and severity-balanced samples.

\begin{table}[h]
\centering
\caption{Distribution of Injury Severity}
\label{tab:injury_severity_new}
\begin{tabular}{|l|r|r|}
\hline
\textbf{Severity} & \textbf{Count} & \textbf{\%} \\ \hline
No Apparent Injury & 490,420 & 78.54\% \\\hline
Possible Injury    & 60,043   & 9.62\% \\\hline
Minor Injury       & 56,830   & 9.10\%  \\\hline
Serious Injury     & 13,811    & 2.21\%  \\\hline
Fatal              & 3,288    & 0.53\%  \\ \hline
\end{tabular}
\end{table}

% \textbf{Data-count verification required before submission.} The manuscript reports 10,130 fatal crashes in the full inventory but later states that 3,288 fatal crashes were available for severity-balanced sampling. It also reports 2,089,634 crashes in the source inventory and 2,069,331 records in the severity-map description. The final text should distinguish the original inventory, the model-eligible analytical dataset, and the subset with valid geographic coordinates, and should report the exclusion reasons at each stage.

\subsection{Model Training}
\subsubsection{Data Sampling and Representativeness}
The three sampling strategies emphasize different objectives. Random sampling approximates the statewide severity distribution, county-based sampling emphasizes geographic coverage, and severity-balanced sampling increases representation of rare outcomes. All experiments use the same model architecture and split ratio. However, each strategy produces a different test subset and class distribution. Accordingly, the reported results describe performance within each strategy-specific evaluation set; direct numerical differences among strategies should not be interpreted as estimates of the effect of training sampling alone.

\subsubsection{Sampling Techniques}

Three model datasets were constructed, each emphasizing a different sampling objective:

\begin{itemize}
    \item \textbf{Random sampling:} 20,000 crashes were selected without explicit county or severity quotas.
    \item \textbf{County-based sampling:} crashes were sampled using county-level targets to increase geographic representation across Tennessee.
    \item \textbf{Severity-balanced sampling:} the analytical sample increased the representation of rare severity classes. The in-sample pool contained 4,000 records for each nonfatal class and 3,288 fatal records. A separate balanced pool contained 3,288 records per class and was divided into training, validation, and unseen test subsets using a 70:15:15 ratio.
\end{itemize}

Across all three strategies, records were converted into the same textualized prompt format. Each strategy-specific dataset was divided into training, validation, and unseen test subsets using a 70:15:15 ratio. Unseen test records were excluded from model training and checkpoint selection and were used only for final evaluation.

\textbf{Two evaluation settings are reported. \emph{In-sample performance} was obtained from an experiment in which the model was trained and evaluated on the full sampled dataset. \emph{Unseen test-data performance} was obtained from a separate experiment in which the sampled dataset was divided into training, validation, and test subsets.} The model was trained on the training subset, and the validation subset was used to select the best checkpoint, defined as a saved model state at a particular point during training. The selected checkpoint was then evaluated on the unseen test subset. Because these results come from separate training runs, their differences are descriptive and should not be interpreted as a conventional training-to-test generalization gap. The unseen test subsets also differ among the three sampling strategies, so cross-strategy comparisons are descriptive rather than controlled comparisons on a common test set.

\begin{table*}[htbp]
  \centering
  \caption{Crash record counts by severity class under the in-sample and unseen-test evaluation settings.}
  \label{tab:record-counts-full}

  \resizebox{\textwidth}{!}{%
  \begin{tabular}{llrrrrr}
    \toprule
    & &
    \multicolumn{1}{c}{\textbf{In-Sample Evaluation}} &
    \multicolumn{4}{c}{\textbf{Unseen-Test Evaluation}} \\
    \cmidrule(lr){3-3}
    \cmidrule(lr){4-7}

    \textbf{Sampling Method} &
    \textbf{Class} &
    \textbf{Full Dataset} &
    \textbf{Train} &
    \textbf{Validation} &
    \textbf{Unseen Test} &
    \textbf{Total} \\

    \midrule
    Random & No Apparent Injury & 15,715 & 10,995 & 2,357 & 2,363 & 15,715 \\
           & Possible Injury    & 1,938  & 1,360  & 288   & 290   & 1,938  \\
           & Minor Injury       & 1,809  & 1,274  & 264   & 271   & 1,809  \\
           & Serious Injury     & 439    & 302    & 71    & 66    & 439    \\
           & Fatal              & 99     & 69     & 20    & 10    & 99     \\
           & \textbf{Total}     & \textbf{20,000} & \textbf{14,000} &
             \textbf{3,000} & \textbf{3,000} & \textbf{20,000} \\

    \midrule
    County & No Apparent Injury & 15,322 & 10,682 & 2,343 & 2,297 & 15,322 \\
           & Possible Injury    & 1,346  & 915    & 227   & 204   & 1,346  \\
           & Minor Injury       & 2,262  & 1,621  & 332   & 309   & 2,262  \\
           & Serious Injury     & 785    & 571    & 106   & 108   & 785    \\
           & Fatal              & 197    & 149    & 26    & 22    & 197    \\
           & \textbf{Total}     & \textbf{19,912} & \textbf{13,938} &
             \textbf{3,034} & \textbf{2,940} & \textbf{19,912} \\

    \midrule
    Severity-Balanced & No Apparent Injury & 4,000 & 2,302 & 493 & 493 & 3,288 \\
           & Possible Injury    & 4,000 & 2,302 & 493 & 493 & 3,288 \\
           & Minor Injury       & 4,000 & 2,302 & 493 & 493 & 3,288 \\
           & Serious Injury     & 4,000 & 2,302 & 493 & 493 & 3,288 \\
           & Fatal              & 3,288 & 2,302 & 493 & 493 & 3,288 \\
           & \textbf{Total}     & \textbf{19,288} & \textbf{11,510} &
             \textbf{2,465} & \textbf{2,465} & \textbf{16,440} \\

    \bottomrule
  \end{tabular}%
  }

  \vspace{2pt}
  \begin{minipage}{\textwidth}
  \footnotesize
  \textit{Note:} In-sample evaluation trains and evaluates the model on the
  full sampled dataset. Unseen-test evaluation is a separate experiment in
  which the dataset is divided into training, validation, and unseen test
  subsets.
  The large reduction in severity-based unseen-test evaluation is because only 3,288 fatal cases are available; therefore, we reduced the sample size to balance the classes.
  \end{minipage}
\end{table*}
Table~\ref{tab:record-counts-full} reports the severity composition of the in-sample pools and the strategy-specific training, validation, and unseen test subsets. The random and county-based unseen test subsets contain only 10 and 22 fatal crashes, respectively. Their fatal-class F1-scores can therefore change substantially based on a small number of predictions and should be interpreted cautiously. The severity-balanced unseen test subset contains 493 fatal records, producing a more stable estimate within that balanced evaluation population. However, this difference in test-set composition also prevents a direct comparison of the three fatal-class F1-scores as though they were calculated on the same population.

\subsubsection{Model Description}

The base model is Llama~3.1~8B. Updating all model parameters would require substantial memory and computation, so the study uses low-rank adaptation (LoRA). LoRA keeps the pretrained weight matrices fixed and learns lower-rank update matrices within selected model projections. This reduces the number of trainable parameters while adapting model behavior to the crash-severity task. Freezing pretrained weights limits direct modification of those parameters, but it does not guarantee that all pretrained capabilities remain unchanged after the adapters are applied.

LoRA adapters are applied to seven linear projections in each of the 32 decoder layers: four attention projections and three feedforward projections, resulting in 224 adapted projections. Each adapter uses rank 8 and scaling factor 16. The base model is loaded with 4-bit quantization. The input and output embedding layers are also trained because five new severity tokens are added to the tokenizer. These tokens represent no apparent injury, possible injury, minor injury, serious injury, and fatal, framing prediction as the generation of one constrained target token rather than open-ended text.

The released SafeTraffic codebase was developed around Illinois and Washington data. The present study retains the model-facing workflow while replacing source-specific preprocessing with Tennessee field mapping and textualization. The resulting experiments evaluate adaptation to Tennessee records; they do not test whether a model trained in one state predicts crashes accurately in another state.

\subsubsection{Training, Validation, and Test Design}
Model development uses training, validation, and unseen test subsets. The training subset is used to update the trainable model parameters. During training, a checkpoint is saved every 25 steps. The validation subset is evaluated at the same interval, and the checkpoint with the highest validation F1-score is selected as the final model. The validation subset does not directly update model parameters but influences model selection. The selected checkpoint is then evaluated on the unseen test subset, which is not used for parameter updating or checkpoint selection. It therefore provides an independent performance estimate for records from the same strategy-specific sampling pool. The term unseen test does not imply temporal, geographic, or cross-state external validation.

% Model development uses training, validation, and unseen test subsets. The training subset updates the trainable parameters. During training, a checkpoint, representing the saved model state at a particular training step, was created every 25 steps. The validation subset was evaluated at the same interval, and the checkpoint with the highest validation F1-score was selected as the final model. The validation subset did not directly update model parameters, but it influenced model selection. The selected checkpoint was then evaluated on the unseen test subset, which was not used for parameter updating or checkpoint selection.
% The validation subset is evaluated every 25 steps and is used to select the checkpoint with the best validation F1-score; it does not directly update model parameters but does influence model selection. The unseen test subset is evaluated after checkpoint selection and is not used for parameter updating or checkpoint selection. It therefore provides an independent estimate for records from the same strategy-specific sampling pool. The term unseen test does not imply temporal, geographic, or cross-state external validation.

The experiments use a 70:15:15 split, consistent with the anchor SafeTraffic Copilot study. The exact split sizes differ by sampling strategy and are reported in Table~\ref{tab:data-splits}. For the 20,000-record random sample, 14,000 records were used for training, 3,000 for validation, and 3,000 for testing on unseen records.

\begin{table}[h]
  \centering
  \caption{Train/validation/test split sizes}
  \label{tab:data-splits}
  \begin{tabular}{|l|r|r|r|r|}
    \hline
    \textbf{Sampling Type} & \textbf{Train} & \textbf{Val} & \textbf{Test} & \textbf{Total} \\
    \hline
    Random    & 14,000 & 3,000 & 3,000 & 20,000 \\\hline
    County    & 13,938 & 3,034 & 2,940 & 19,912 \\\hline
    Severity  & 11,510 & 2,465 & 2,465 & 16,440 \\
    \hline
  \end{tabular}
\end{table}

\subsection{Computational Resources}

All model trainings were conducted on Perlmutter supercomputer at National Energy Research Scientific Computing Center (NERSC) using one AMD EPYC 7763 CPU and two to four NVIDIA A100 GPUs with 40 or 80 GB of memory. Training was distributed with DeepSpeed ZeRO Stage~2, and two to four GPUs were allocated depending on resource availability. All experiments used Python~3.9. The installed CUDA driver version was 12.9, while PyTorch was compiled against CUDA~12.1; DeepSpeed was configured with \texttt{DS\_SKIP\_CUDA\_CHECK=1}. No failure was attributed to this version difference during the completed runs. Training used bfloat16 precision, a 4-bit-quantized Llama~3.1~8B base model, and LoRA adapters with rank 8 and alpha 16.

All sampling strategies used a maximum sequence length of 4{,}096 tokens, a learning rate of $3\times10^{-4}$, and random seed 42. Training targeted two epochs, with 20 warmup steps and evaluation and checkpointing every 25 steps. The per-device batch size was 2, and the effective batch size was 64 for completed experiments through strategy-specific combinations of GPU count and gradient accumulation. 

In-sample evaluations used two 40 GB A100 GPUs for each sampling strategy. The random and county-based unseen-test experiments used two 80 GB A100 GPUs, while the severity-balanced unseen-test experiment used four 80 GB A100 GPUs.

\section{Experiments and Results}
\begin{table*}[h]
  \centering
% PREVIOUS TEXT:   \caption{Results overview: in-sample vs.\ out-of-sample}
  \caption{Results overview: in-sample vs.\ strategy-specific unseen test-data performance}
  \label{tab:results-overview}
  \resizebox{\textwidth}{!}{%
  \begin{tabular}{|l|c|c|c|c|c|c|}
    \hline
    & \multicolumn{2}{|c|}{\textbf{Accuracy}} & \multicolumn{2}{|c|}{\textbf{F1-score (weighted)}} & & \\\hline
% PREVIOUS TEXT:     \textbf{Sampling Method} & \textbf{In-Sample} & \textbf{Out-of-Sample} & \textbf{In-Sample} & \textbf{Out-of-Sample} & \textbf{$\Delta$ Acc.} & \textbf{$\Delta$ F1} \\
    \textbf{Sampling Method} & \textbf{In-Sample} & \textbf{Unseen Test Data} & \textbf{In-Sample} & \textbf{Unseen Test Data} & \textbf{$\Delta$ Acc.} & \textbf{$\Delta$ F1} \\
    \hline
    Random   & 90.5\% & 81.4\% & 91.0\% & 81.8\% & $-$9.1\%  & $-$9.2\%  \\\hline
    County   & 93.5\% & 79.1\% & 93.8\% & 80.1\% & $-$14.4\% & $-$13.7\% \\\hline
    Severity & 65.0\% & 58.0\% & 64.9\% & 57.7\% & $-$7.0\%  & $-$7.2\%  \\
    \hline
  \end{tabular}%
  }
\end{table*}

{\color{orange}

}

Table~\ref{tab:results-overview} shows lower unseen test-data performance than in-sample performance for all three strategy-specific experiments. On their respective unseen test subsets, the random model achieved 81.4\% accuracy and 81.8\% weighted F1, the county-based model achieved 79.1\% accuracy and 80.1\% weighted F1, and the severity-balanced model achieved 58.0\% accuracy and 57.7\% weighted F1. These values should not be ranked as a controlled comparison because the unseen test records and class distributions differ across strategies. Within the random and county-based evaluation sets, aggregate measures are also strongly influenced by the dominant no-apparent-injury class.

\subsection{Strategy-Specific Performance Profiles}

\begin{figure*}[htbp]
    \centering
    \includegraphics[width=\textwidth]{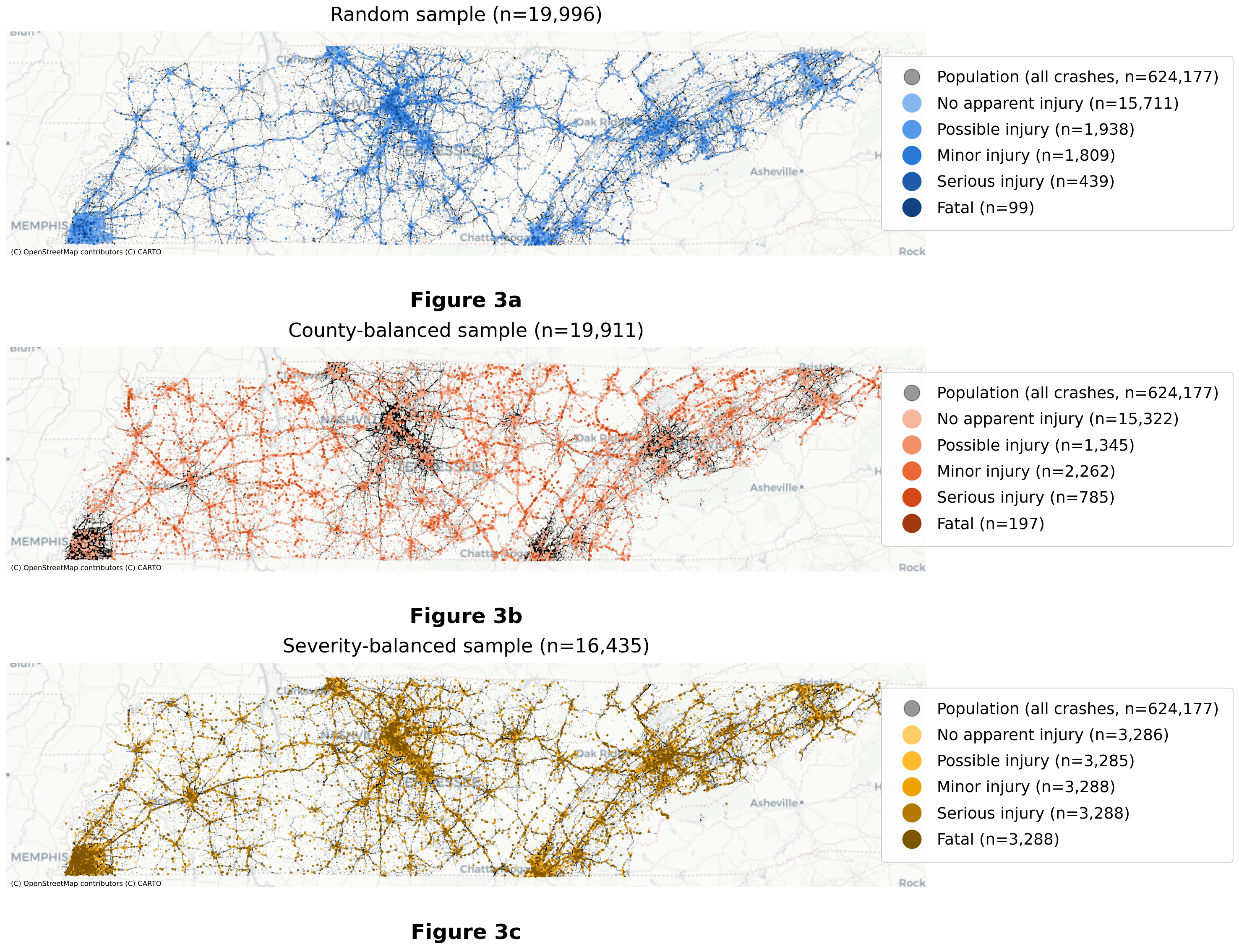}
    \caption{Spatial distributions of the three sampling strategies.}
    \Description{Three Tennessee maps compare the spatial patterns produced by random, county-based, and severity-balanced crash sampling.}
    \label{fig:all-heatmap}
    {\footnotesize\textit{Note:} Plotted counts are slightly lower than the sample sizes reported in the results tables, as records missing geographic coordinates could not be mapped and were excluded from this figure only.}
\end{figure*}

\subsubsection{Random Sampling}

Figure \ref{fig:all-heatmap}a represents the spatial distribution of the random 20000 crashes selected from the database of Tennessee crashes. Each point represents a single crash, colored by severity level. Lighter shades indicate no apparent injury, and darker shades indicate increasing severity up to fatal crashes. The sample closely mirrors the geographic spread of the full population, concentrating along major highway corridors and urban centers such as Nashville, Memphis, Knoxville, and Chattanooga.

\begin{table*}[htbp]
  \centering

  \caption{Random sampling: in-sample vs.\ unseen test data F1-score by class.}
  \label{tab:random-compare}
  \begin{threeparttable}
  \begin{tabular}{|l|c|c|c|c|}
    \hline

    \textbf{Class} & \textbf{F1 (In-Sample)} & \textbf{F1 (Unseen Test)} & \textbf{$\Delta$ F1} & \textbf{Support (In-Sample / Unseen Test)} \\
    \hline
    No Apparent Injury & 94.4\% & 92.4\% & $-$2.0 & 15,715 / 2,363 \\\hline
    Possible Injury & 71.2\% & 38.7\% & \textbf{$-$32.5} & 1,938 / 290 \\\hline
    Minor Injury & 78.2\% & 49.2\% & \textbf{$-$29.0} & 1,809 / 271 \\\hline
    Serious Injury & 99.0\% & 36.4\% & \textbf{$-$62.6} & 439 / 66 \\\hline
    Fatal & 98.5\% & 17.4\% & \textbf{$-$81.1} & 99 / 10 \\
    \hline
    Accuracy & 90.5\% & 81.4\% & $-$9.1 & 20,000 / 3,000 \\\hline
    Macro avg F1 & 88.3\% & 46.8\% & \textbf{$-$41.5} & 20,000 / 3,000 \\\hline
    Weighted avg F1 & 91.0\% & 81.8\% & $-$9.2 & 20,000 / 3,000 \\
    \hline
  \end{tabular}
  % \begin{tablenotes}
  %   \footnotesize

  %   % \item \textcolor{green!50!black}{\textit{Note.} Bold $\Delta$ F1 indicates a drop of 20 points or more between in-sample and unseen test-data performance.}
  % \end{tablenotes}
  \end{threeparttable}
\end{table*}

Table~\ref{tab:random-compare} shows strong unseen test-data performance for no apparent injury, with an F1-score of 92.4\%, but substantially lower F1-scores for possible injury (38.7\%), minor injury (49.2\%), serious injury (36.4\%), and fatal crashes (17.4\%). Consequently, the weighted F1 on unseen test data remained high at 81.8\%, while macro F1 was 46.8\%. The large differences between in-sample and unseen test-data performance for serious and fatal crashes indicate that the fitted model did not retain comparable performance on unseen records from the random-sampling pool. The fatal estimate is based on only 10 unseen test cases and is highly sensitive to individual predictions.

\subsubsection{County-Based Sampling}

Figure~\ref{fig:all-heatmap}b represents the sample that was taken based on counties, overlaid on the full population of Tennessee crash records. Each point represents a single crash, colored by severity level. This strategy draws a fixed target of 210 crashes from every county, with one exception where only 172 crashes were available. As a result, coverage extends more evenly into rural and lower-density counties than under random sampling. This comes at the cost of lighter representation in the state's highest-crash-density urban corridors.

\begin{table*}[htbp]
  \centering
% PREVIOUS TEXT:   \caption{County-Stratified Sampling: In-Sample vs.\ Out-of-Sample F1-score by class.}
  \caption{County-based sampling: in-sample vs.\ unseen test data F1-score by class.}
  \label{tab:county-compare}
  \begin{threeparttable}
  \begin{tabular}{|l|c|c|c|c|}
    \hline
    
    \textbf{Class} & \textbf{F1 (In-Sample)} & \textbf{F1 (Unseen Test)} & \textbf{$\Delta$ F1} & \textbf{Support (In-Sample / Unseen Test)} \\
    \hline
    No Apparent Injury & 96.2\% & 92.2\% & $-$4.0 & 15,322 / 2,297 \\\hline
    Possible Injury & 78.5\% & 22.1\% & \textbf{$-$56.4} & 1,346 / 204 \\\hline
    Minor Injury & 85.5\% & 43.6\% & \textbf{$-$41.9} & 2,262 / 309 \\\hline
    Serious Injury & 97.6\% & 46.9\% & \textbf{$-$50.7} & 785 / 108 \\\hline
    Fatal & 98.2\% & 26.7\% & \textbf{$-$71.5} & 197 / 22 \\
    \hline
    Accuracy & 93.5\% & 79.1\% & $-$14.4 & 19,912 / 2,940 \\\hline
    Macro avg F1 & 91.2\% & 46.3\% & \textbf{$-$44.9} & 19,912 / 2,940 \\\hline
    Weighted avg F1 & 93.8\% & 80.1\% & $-$13.7 & 19,912 / 2,940 \\
    \hline
  \end{tabular}
  % \begin{tablenotes}
  %   \footnotesize

  %   \item \textcolor{green!50!black}{\textit{Note.} Bold $\Delta$ F1 indicates a drop of 20 points or more between in-sample and unseen test-data performance.}
  % \end{tablenotes}
  \end{threeparttable}
\end{table*}

Table~\ref{tab:county-compare} shows that the county-based model achieved 92.2\% unseen test-data F1 for no apparent injury, compared with 22.1\% for possible injury, 43.6\% for minor injury, 46.9\% for serious injury, and 26.7\% for fatal crashes. Its unseen test-data macro F1 was 46.3\%, while weighted F1 was 80.1\%, again showing the influence of the dominant class on the aggregate result. The broader county coverage in the sample did not remove severity imbalance within this experiment. The fatal-class result is based on 22 unseen test cases and remains unstable without repeated sampling or confidence intervals.

\subsubsection{Severity-Based Sampling}
Figure~\ref{fig:all-heatmap}c represents the data sampled based on severity overlaid on the full population of Tennessee crash records. Each severity class is represented in approximately equal numbers (ranging from 3{,}285 to 3{,}288 crashes per class), rather than in proportion to its natural frequency in the population. As a result, rare severity outcomes such as serious injury and fatal crashes appear far more densely on this map than they do in the true population distribution, reflecting the deliberate class-balancing applied by this strategy rather than the actual geographic concentration of severe crashes

\begin{table*}[h]
  \centering

  \caption{Severity-balanced sampling: in-sample vs.\ unseen test data F1-score by class.}
  \label{tab:severity-compare}
  \begin{threeparttable}
  \begin{tabular}{|l|c|c|c|c|}
    \hline

    \textbf{Class} & \textbf{F1 (In-Sample)} & \textbf{F1 (Unseen Test)} & \textbf{$\Delta$ F1} & \textbf{Support (In-Sample / Unseen Test)} \\
    \hline
    No Apparent Injury & 81.6\% & 76.3\% & $-$5.3 & 4,000 / 493 \\\hline
    Possible Injury & 57.0\% & 48.8\% & $-$8.2 & 4,000 / 493 \\\hline
    Minor Injury & 57.2\% & 51.2\% & $-$6.0 & 4,000 / 493 \\\hline
    Serious Injury & 57.9\% & 46.4\% & $-$11.5 & 4,000 / 493 \\\hline
    Fatal & 71.9\% & 65.9\% & $-$6.0 & 3,288 / 493 \\
    \hline
    Accuracy & 65.0\% & 58.0\% & $-$7.0 & 19,288 / 2,465 \\\hline
    Macro avg F1 & 65.1\% & 57.7\% & $-$7.4 & 19,288 / 2,465 \\\hline
    Weighted avg F1 & 64.9\% & 57.7\% & $-$7.2 & 19,288 / 2,465 \\
    \hline
  \end{tabular}
  % \begin{tablenotes}
  %   \footnotesize

  %   \item \textcolor{green!50!black}{\textit{Note.} Bold $\Delta$ F1 indicates a drop of 20 points or more between in-sample and unseen test-data performance.}
  % \end{tablenotes}
  \end{threeparttable}
\end{table*}

Table~\ref{tab:severity-compare} shows unseen test-data F1-scores ranging from 46.4\% to 76.3\% across the five equally supported classes. The differences between in-sample and unseen test-data performance were smaller than those observed in the random and county-based experiments. Fatal-crash F1 was 65.9\% within the balanced unseen test subset. The macro and weighted F1-scores were both 57.7\% because every unseen test class had equal support. Comparisons with the random and county-based fatal F1-scores should remain descriptive because those estimates were calculated on different test populations with substantially fewer fatal records.

\section{Discussion}
The experiments show that aggregate and class-level conclusions can differ substantially within each strategy-specific evaluation set. The random and county-based models achieved unseen-test-data weighted F1-scores near 80\%, but their macro F1-scores remained below 47\% and their fatal-crash F1-scores remained below 27\%. These results indicate that high weighted performance on naturally imbalanced subsets can be driven largely by the dominant no-apparent-injury class. For safety-oriented use, macro F1, class-level precision and recall, and the number of test examples per class are therefore necessary complements to aggregate accuracy and weighted F1.

The severity-balanced experiment produced a different performance profile. Its unseen test subset contained equal class support, and the model achieved a fatal-crash F1-score of 65.9\% with smaller differences between in-sample and unseen test-data performance across classes. Because this model was evaluated on a different and artificially balanced test population, its aggregate scores cannot be directly ranked against those from the random and county-based test subsets. The result supports the narrower conclusion that increasing rare-class representation can produce a more balanced class-level profile within the evaluated sample. It does not establish that severity-balanced sampling is universally superior for operational deployment on the natural statewide distribution.

The county-based experiment increased geographic representation but retained substantial severity imbalance. Its results illustrate that geographic and severity representation are separate design dimensions. Future work could evaluate hybrid sampling or class-sensitive objectives while holding the test population fixed. Such an experiment would be required to isolate the effect of the training sampling strategy.

The scope of transfer demonstrated here is workflow adaptation. The released Illinois and Washington preprocessing and model-facing interface were reproduced and adapted to Tennessee records. The predictive results, however, are entirely within Tennessee and use strategy-specific random splits. They do not establish temporal transfer, geographic transfer to unseen counties, or cross-state predictive transfer. The contribution is therefore an operational adaptation of an LLM crash-severity pipeline and a descriptive analysis of how reported performance profiles vary across sampling and evaluation designs.

\section{Conclusion}\label{sec:conclusion}

This study reproduced and adapted an LLM-based crash-severity pipeline to a three-year Tennessee inventory and documented performance for random, county-based, and severity-balanced samples. On their respective unseen test subsets, the random and county-based models achieved weighted F1-scores of 81.8\% and 80.1\%, but fatal-crash F1-scores of only 17.4\% and 26.7\%. The severity-balanced experiment achieved weighted and macro F1-scores of 57.7\% and a fatal-crash F1-score of 65.9\%. Because the test subsets differed in both records and class composition, these values describe different evaluation populations and should not be interpreted as a controlled ranking of sampling strategies. The consistent practical lesson is that aggregate metrics alone are insufficient for a strongly imbalanced safety task. Evaluations should report macro and class-level measures, test support, and the composition of the evaluation population alongside accuracy and weighted F1.

\section{Limitations and Future Work}

The study evaluates one base model, one LoRA configuration, one random seed, and one state inventory. Each sampling strategy uses a different unseen test subset, so the observed cross-strategy differences combine the effects of training-sample composition and test-population composition. The experiments therefore do not isolate the causal effect of sampling strategy. In-sample results are also calculated on records used for model fitting and should be interpreted only as measures of training fit, not independent predictive performance. In addition, the random and county-based unseen test subsets contain only 10 and 22 fatal crashes, respectively, making their fatal-class estimates unstable.

The current work also does not test temporal, unseen-county, or cross-state transfer. The input construction should be audited to verify that fields or narrative phrases that directly reveal final injury severity are excluded when the intended task is early severity prediction. Records without valid geographic coordinates were excluded from the spatial visualizations but retained in the analytical datasets, resulting in slightly different mapped and analytical sample counts.

Future work should evaluate all trained models on the same untouched naturally distributed test set and, when useful, on a second common severity-balanced diagnostic set. Repeated experiments, bootstrap confidence intervals, temporally separated testing, conventional statistical and machine-learning baselines, structured-only and narrative-only ablations, calibration, and sentence-level attribution would further establish whether the observed rare-class performance is reliable and operationally meaningful.

% ---------- Acknowledgments ----------
\begin{acks}
The authors thank the Tennessee Department of Transportation for providing the statewide crash data used in this study. The authors also acknowledge the developers of SafeTraffic Copilot for making the source code and supporting data publicly available. 

The authors used OpenAI to assist with language editing, document organization, and preliminary literature synthesis. All model-generated content was reviewed and revised by the authors, who take full responsibility for the final manuscript, including the accuracy of its citations, technical statements, analyses, interpretations, and conclusions. 

This research used resources of the National Energy Research Scientific Computing Center (NERSC), a U.S. Department of Energy Office of Science User Facility supported by the Office of Science of the U.S. Department of Energy under Contract No. DE-AC02-05CH11231.
\end{acks}
% ---------- Author Contributions ----------
% Consider using the Contributor Role Taxonomy (CRediT) https://credit.niso.org/ when stating author contributions
\section*{AUTHOR CONTRIBUTIONS}
The authors confirm contribution to the paper as follows: 
study conception and design: A. Saroj, P. Govindu, B. Sharma, U. Ahmed; 
data collection: A. Saroj; 
analysis and interpretation of results: A. Saroj, P. Govindu, B. Sharma, U. Ahmed; 
draft manuscript preparation:  A.Saroj, P. Govindu. 
All authors reviewed the results and approved the final version of the manuscript.

% ---------- Declaration of COI ----------
% \section*{DECLARATION OF CONFLICTING INTERESTS}
% % Choose one of the following statements:
% All authors declare no potential conflicts of interest with respect to the research, authorship, and publication of this article.

% % OR
%The authors declared the following potential conflicts of interest with respect to the research, authorship, and/or publication of this article: \emph{[insert text here]}.

% % OR
% The authors declared no potential conflicts of interest with respect to the research, authorship, and/or publication of this article.

\section*{FUNDING}
% Choose one of the following statements:
% The authors disclosed receipt of the following financial support 
% for the research, authorship, and/or publication of this article: 
% This research was supported by \emph{[funding agency]} (grant no.~\emph{xxxxx}).

% % OR
The authors disclosed no financial support for the research, authorship, and/or publication of this article.
\section*{Legal Disclaimer}
This material was prepared as an account of work sponsored by an agency of the United States Government. Neither the United States Government nor any agency thereof, nor any of its employees, makes any warranty, express or implied, or assumes any legal liability or responsibility for the accuracy, completeness, or usefulness of any information, apparatus, product, or process disclosed, or represents that its use would not infringe privately owned rights. Reference herein to any specific commercial product, process, or service by trade name, trademark, manufacturer, or otherwise does not necessarily constitute or imply its endorsement, recommendation, or favoring by the United States Government or any agency thereof. The views and opinions of authors expressed herein do not necessarily state or reflect those of the United States Government or any agency thereof.

% ----------------------------------------------------------------
% REFERENCES
% ----------------------------------------------------------------
\bibliographystyle{ACM-Reference-Format}
\bibliography{references}

\end{document}